\documentclass[prl,twocolumn,aps,superscriptaddress,showpacs]{revtex4}
\usepackage{epsfig}
\usepackage{graphicx}
\usepackage{amsmath}
\usepackage{amscd}

\newcommand{\beq}{\begin{eqnarray}}
\newcommand{\eeq}{\end{eqnarray}}
\def\be{\begin{equation}}
\def\ee{\end{equation}}
\def\ba{\begin{eqnarray}}
\def\ea{\end{eqnarray}}

\begin{document}
\title
{All-optical Imprinting of Geometric Phases onto Matter Waves}
\author{Qi Zhang}
\affiliation{Centre of Quantum Technologies and Department of
Physics, National University of Singapore, 117543, Singapore}

\author{Jiangbin Gong}
\affiliation{Department of Physics and Centre for Computational
Science and Engineering, National University of Singapore, 117542,
Singapore}
 \affiliation{NUS Graduate School for Integrative Sciences
and Engineering, Singapore 117597, Republic of Singapore}

\author{C.H. Oh}
\affiliation{Centre of Quantum Technologies and Department of
Physics, National University of Singapore, 117543, Singapore}

\date{\today}
%
\date{\today}
\begin{abstract}
Traditional optical phase imprinting of matter waves is of a
dynamical nature. In this paper we show that both Abelian and
non-Abelian geometric phases can be optically imprinted onto matter
waves, yielding a number of interesting phenomena such as wavepacket
re-directing and wavepacket splitting. In addition to their
fundamental interest, our results open up new opportunities for
robust optical control of matter waves.
\end{abstract}
\pacs{03.65.Vf, 03.75.-b, 32.80.Qk} \maketitle

When the Hamiltonian of a quantum system adiabatically changes along
a path in its parameter space, a non-degenerate (degenerate)
eigenstate will acquire, on top of an obvious dynamical phase,  an
Abelian (non-Abelian) geometric phase
\cite{Berry1984PRS,Wilczek1984PRL}. The discovery of geometric
phases has motivated many proposals on built-in fault-tolerant
quantum
logical operations \cite{zanardi1999}.  
Geometric phases also provide an elegant framework to understand the
coupling between translational motion and internal degrees of
freedom.  In particular,  a slowly moving system subject to an
external field naturally causes adiabatic changes in its internal
Hamiltonian, giving rise to an Abelian or non-Abelian vector
potential for the translational motion
\cite{Berry1993,Littlejohn1993PRA,niu}.
Interestingly, one can also engineer laser-matter interaction to
synthesize effective vector potentials from geometric phases
\cite{dum,Juzeliunas2004PRL,Juzeliunas2005PRA,Ruseckas2005PRL}.

One implicit assumption in geometric phase-based quantum gate
studies is that the qubit systems are frozen in space. By contrast,
here we examine the geometric phases acquired by a system coherently
delocalized in space.  In particular, we propose to optically
imprint geometric phases at each local space point onto delocalized
matter waves of ultracold systems. We show that the imprinted
geometric phases can alter the ensuing matter-wave propagation
dramatically, thus establishing another interesting and potentially
powerful means of optical control.  Because here the geometric
phases are not induced by the translational motion itself, but by
active manipulation of laser-matter interaction, the imprinting of
geometric phases onto matter waves results in a new type of coupling
between internal and translational motions, which can be implemented
experimentally.

In generating matter-wave vortices with time-dependent magnetic
fields, some early work did take advantage of a spatially-dependent
Abelian Berry phase \cite{vortex}.  Our work however represents a
major extension from the vortex context, by demonstrating the
imprinting of both Abelian and non-Abelian geometric phases with
{\it all-optical} methods.  We shall show that imprinting geometric
phases may re-direct atom wavepackets, split a wavepacket, or create
interesting matter wave patterns.
The all-optical geometric phase imprinting is insensitive to the
dynamical details. Hence it is superior to conventional optical
phase imprinting approach to matter wave engineering
\cite{dobrek1999,science2000}, where the imprinted phase is
proportional to the duration of optical imprinting and to the
magnitude of an optical potential.

Consider first a two-level atom in a plane-wave laser field. In the
rotating wave approximation (RWA), the internal Hamiltonian is given
by $ H_{\text{RWA},2}= (\Delta/2)(|1\rangle\langle
1|-|2\rangle\langle 2|) +
(\Omega_{0}/2)(e^{-i\mathbf{k}^{l}\cdot\mathbf{R}} |1\rangle\langle
2| +  e^{i\mathbf{k}^{l}\cdot\mathbf{R}} |2\rangle \langle 1|)$,
where $|1\rangle$ and $|2\rangle$ are the two internal states of the
atom, $\Omega_{0}$ is assumed to be real, ${\bf k}^{l}$ is the
wavevector of the laser field, $\Delta$ is the detuning from
one-photon resonance, and ${\bf R}$ the coordinate of the atom.
In the $(|1\rangle, |2\rangle)$ representation
$H_{\text{RWA},2}$ has the eigenstate
\begin{eqnarray} \label{eigenstate1}
|\phi^{+}\rangle&=&\frac{1}{\chi(\Delta)}\left(\begin{array}{c}
\sqrt{\Omega_{0}^{2}+\Delta^{2}}+\Delta\\
\Omega_{0}e^{i\mathbf{k}^{l}\cdot\mathbf{R}}
\end{array}
 \right),
%
\end{eqnarray}
and an analogous eigenstate $|\phi^{-}\rangle$ orthogonal to
$|\phi^{+}\rangle$, where
$\chi(\Delta)=\sqrt{\Omega_{0}^{2}+(\sqrt{\Omega_{0}^{2}+\Delta^{2}}+\Delta)^{2}}$
is the normalization factor.  Note that these internal eigenstates
vary with the coordinate ${\bf R}$.  Within the adiabatic
approximation and in the absence of a trapping potential, the
overall eigenstate (internal plus translational) for the system,
denoted $|\Psi^{\pm}\rangle$, is given by
$|\Psi^{\pm}\rangle=|\phi^{\pm}\rangle
e^{i\mathbf{k}\cdot\mathbf{R}}$, where ${\bf k}$ is a wavevector
associated with the translational motion of the atom.

We begin with a test case that is well-known in laser chirping
control \cite{chirping} (which realizes a Landau-Zener process).
Here the detuning parameter $\Delta$ is adiabatically varied from
$\Delta_1>0$ to $\Delta_{2}<0$, with $|\Delta_{1}/\Omega_0| \gg 1$
and $|\Delta_{2}/\Omega_0| \gg 1$. Accordingly ${\bf k}^{l}$ changes
from ${\bf k}^{l}_{1}$ to ${\bf k}^{l}_{2}$, but with a fixed
direction. The initial state is assumed to be
$|\Psi^{+}(\Delta=\Delta_1)\rangle\approx (1,0)^{\text{T}} e^{i{\bf
k}\cdot {\bf R}}$. Assuming that the translational motion during the
chirping process is negligible, which can be a good approximation
for ultracold atoms, one finds the following geometric phase for an
arbitrary ${\bf R}$,
\begin{eqnarray}
\beta(\mathbf{R})&=&i \int_{\Delta_{1}}^{\Delta_{2}}
\langle \phi^{+}|\frac{\partial}{\partial \Delta}|\phi^{+}\rangle d\Delta \nonumber \\
&=& -\frac{\mathbf{R}\cdot\hat{\bf k}^{l}}{\hbar c}
\int_{\Delta_{1}}^{\Delta_{2}} \frac{\Omega_{0}^{2}}{\chi^2(\Delta)}
d \Delta \nonumber \\
&\approx &  -{\bf R}\cdot \left[{\bf k}^{l}_{2}-{\bf
k}^{l}_{r}\right], \label{GP}
\end{eqnarray}
where ${\bf k}^{l}_{r}$ denotes the laser wavevector when $\Delta
=0$ \cite{footnote}, and throughout $\hat{\bf k}$ represents a unit
vector along the vector ${\bf k}$.  Given that the evolving state
remains in the system eigenstate, the dynamical phase
 is evidently ${\bf R}$-independent. Imprinting the
geometric phase $\beta({\bf R})$ onto the adiabatic state
$|\phi^{+}\rangle$ in Eq. (\ref{eigenstate1}) and neglecting the
overall dynamical phase, one obtains the final state
\begin{eqnarray}
&&|\Psi^{+}_{f}(\Delta_{2})\rangle = \frac{ e^{i[\beta({\bf R})+{\bf
k}\cdot{\bf R}]}}{\chi(\Delta_2)}\left(\begin{array}{c}
\sqrt{\Omega_{0}^{2}+\Delta^{2}_{2}}+\Delta_{2}\\
\Omega_{0}e^{i\mathbf{k}^{l}_{2}\cdot\mathbf{R}}
\end{array} \right)  \nonumber \\
& & \approx  e^{i[\beta({\bf R})+{\bf k}\cdot{\bf R}]}
\left(\begin{array}{c}0 \\
e^{i\mathbf{k}^{l}_{2}\cdot\mathbf{R}}\end{array} \right)
=\left(\begin{array}{c}0 \\
1\end{array} \right)e^{i({\bf k}_r^{l}+{\bf k})\cdot {\bf R}}.
\label{chirpeq}
\end{eqnarray}
Equation (\ref{chirpeq}) shows that the population between the two
levels is adiabatically inverted and the atom momentum increases by
$\hbar{\bf k}^{l}_{r}$. This result is expected, because (i) the
chirping scenario is long known to induce the absorption of one
on-resonance photon and (ii) momentum conservation requires the atom
momentum to increase by $\hbar{\bf k}^{l}_{r}$. But quite
noteworthy, here the obvious consequence of momentum conservation is
manifested as an effect of the spatially-dependent geometric phase
$\beta({\bf R})$. Note also that if we further change $\Delta$ along
a reversed path until it reaches its initial value, then Eq.
(\ref{GP}) gives a zero Berry phase and hence zero momentum change.
This is again correct because the atom releases a photon during the
reversed process.  One can also carry out similar analysis if the
chirped laser field possesses a nonzero orbital angular momentum.
That is, the transfer of orbital angular momentum from photon to the
matter wave (hence vortex generation) can also be interpreted as
geometric phase imprinting.

As a second Abelian example, we examine the system of
$H_{\text{RWA},2}$ for $\Delta=0$ and with
$\hat{\mathbf{k}}^{l}_{r}$ being rotated adiabatically. We assume
below both $\hat{\mathbf{k}}^{l}_{r}$ and ${\bf k}$ are in the
$x$-$z$ plane. Let $ \gamma $ be the angle by which
$\hat{\mathbf{k}}^{l}_{r}$ is rotated clockwise with respect to the
$x$-axis.  Then
${\mathbf{k}}^{l}_{r}(\gamma)=
k^{l}_r\left[\cos(\gamma)\hat{e}_x-\sin(\gamma)\hat{e}_z\right],$
where $k^{l}_{r}\equiv |{\bf k}^{l}_r|$,
$\hat{e}_{x}$ and $\hat{e}_{z}$ are the unit vectors along the
$x$-axis and $z$-axis. For convenience we also define the polar
coordinates $(R,\theta)$ for ${\bf R}$ confined on the $x$-$z$
plane, i.e., $x=R\cos(\theta)$, and $z=R\sin(\theta)$. Then, the
eigenstate in Eq. (\ref{eigenstate1}) for $\Delta=0$ can be
rewritten as
\begin{eqnarray} \label{Rostate}
|\phi^{+}(\gamma)\rangle&=&
\frac{1}{\sqrt{2}}\left(\begin{array}{c}
1\\
e^{i{k}^{l}_{r}R\cos(\theta+\gamma)}
\end{array}
 \right)
\end{eqnarray}
Consider an initial state $|\Psi^{+}\rangle=
|\phi^{+}(\gamma=0)\rangle e^{i{\bf k}\cdot {\bf R}}$. As the
adiabatic parameter $\gamma$ varies from $0$ to $\alpha$, the
acquired geometric phase at an arbitrary local point $(R,\theta)$ is
found to be
\begin{eqnarray}
\beta(R,\theta) &=&i \int_{0}^{\alpha}
\langle\phi^{+}|\frac{\partial}{\partial
\gamma}|\phi^{+}\rangle d\gamma\nonumber \\
&=&\frac{k_{r}^{l}}{2}\left[R\cos(\theta)-R\cos(\theta+\alpha)\right].
\end{eqnarray}
Neglecting the translational motion  during the adiabatic process,
the final state $|\Psi^{+}_{f}(\alpha)\rangle$
 can be obtained by
 imprinting $\beta(R,\theta)$
onto the adiabatic state  $|\phi^{+}(\gamma=\alpha)\rangle e^{i{\bf
k}\cdot {\bf R}}$, yielding
\begin{eqnarray} \label{final}
|\Psi^{+}_{f}(\alpha)\rangle&=& e^{i[\beta(R,\theta)+{\bf
k}\cdot{\bf R}]} |\phi^{+}(\gamma=\alpha)\rangle
\nonumber \\
&=&\frac{1}{\sqrt{2}}\left(\begin{array}{c}
e^{ik^{l}_{r}R\sin(\theta+\frac{\alpha}{2})\sin(\frac{\alpha}{2})}\\
e^{ik^{l}_{r}R\cos(\theta+\frac{\alpha}{2})\cos(\frac{\alpha}{2})}
\end{array}
 \right)e^{i\mathbf{k}\cdot\mathbf{R}}.
\end{eqnarray}
Unlike in the above test case that apparently involves the
absorption of one on-resonance photon, here neither the population
on states $|1\rangle$ and $|2\rangle$ nor the atom mechanical
momentum
changes.   
However, the geometric phases $\beta(R,\theta)$ imprinted onto the
matter wave will induce differences in time evolution if we suddenly
switch off the laser field, after which the matter waves associated
with each of the two components of $|\Psi^{+}_{f}(\alpha)\rangle$
will evolve independently and freely. The momentum in free space for
each component of $|\Psi^{+}_{f}(\alpha)\rangle$ can be found by
taking the spatial derivative of that component. Without loss of
generality we set ${\bf k}=0$. One then finds that upon a sudden
laser field switch-off, the state $|\Psi^{+}_{f}(\alpha)\rangle$
will split into two components: atoms with the internal state
$|2\rangle$ will move in the direction of
$\cos(\alpha/2)\hat{e}_{x}-\sin(\alpha/2)\hat{e}_{z}$ with a
momentum $\hbar k^{l}_{r}\cos(\alpha/2)$, but the atoms on the state
$|1\rangle$ will move in the direction of
$\sin(\alpha/2)\hat{e}_{x}+\cos(\alpha/2)\hat{e}_{z}$ with a
momentum $\hbar k^{l}_{r}\sin(\alpha/2)$.  These motion
characteristics are clearly $\alpha$-dependent and intriguing. In
particular, if there is no geometric phase imprinting, $\alpha=0$,
the $|1\rangle$ component is static and the $|2\rangle$ component
moves along $\hat{e}_x$; but if $\alpha=\pi$, then the $|2\rangle$
component is static and the $|1\rangle$ component moves along
$\hat{e}_{x}$. This could be useful for filtering the internal
states.

We have also carried out numerical wavepacket simulations based on
the Hamiltonian $-\frac{\hbar^{2}\nabla^{2}}{2m}+H_{\text{RWA},2}$
($m$ being the mass of the atom) that undergoes slow changes in its
parameter $\gamma$. In our simulations we replace the plane-wave
factor $\exp(i{\bf k}\cdot {\bf R})$ of $|\Psi^{+}\rangle$ by a
Gaussian wavepacket. Numerical results do confirm our predictions.


We now discuss the imprinting of non-Abelian geometric phases. As in
Refs. \cite{Bergmann,Ruseckas2005PRL,Juzeliunas2008PRL}, we adopt
the so-called ``tripod scheme" of four-level atoms interacting with
three on-resonance laser fields.   The internal state is denoted as
$|n\rangle$, $n=0-3$. Each of the three transitions $|0\rangle
\leftrightarrow |1\rangle$, $|0\rangle\leftrightarrow |2\rangle$,
and $|0\rangle\leftrightarrow |3\rangle$ is coupled by one laser
field.  This coupling scheme can be realized if states $|1\rangle$,
$|2\rangle$, and $|3\rangle$ are degenerate magnetic sub-levels and
the three coupling fields have different polarizations.
Alternatively, states $|1\rangle$, $|2\rangle$, $|3\rangle$ are
non-degenerate and the three coupling fields then carry different
frequencies (same polarization is then allowed). For convenience we
adopt the same configuration as in Ref. \cite{Juzeliunas2008PRL},
where two laser beams are counter-propagating along the $x$-axis and
the third laser beam is along the $z$-axis. The associated internal
Hamiltonian under RWA is given by
$H_{\text{RWA},4}=\sum_{n=1}^{3}\Omega_{n}|0\rangle\langle n| +
h.c.,$
with $\Omega_{1}=\Omega_{0}\sin(\xi)/\sqrt{2}e^{-ik_{r}^{l}x}$,
$\Omega_{2}=\Omega_{0}\sin(\xi)/\sqrt{2}e^{ik_{r}^{l}x}$, and
$\Omega_{3}=\Omega_{0}\cos(\xi)e^{ik_{r}^{l}z}$, where the parameter
$\xi$ is set to satisfy $\cos(\xi)=\sqrt{2}-1$.  Note that even if
the three lasers carry different frequencies, their effective
$k^{l}_{r}$ in the $x$-$z$ plane can still be the same by titling
the laser beams out of the $x$-$z$ plane.

The Hamiltonian $H_{\text{RWA},4}$ has two degenerate and
spatially-dependent dark (null-eigenvalue) states
$|D_{1(2)}\rangle$, which are given by \cite{Juzeliunas2008PRL},
\begin{eqnarray}
|D_1\rangle&=&(|\tilde{1}\rangle-|\tilde{2}\rangle)e^{-i\kappa'
z}/\sqrt{2} \nonumber \\
|D_2\rangle&=&\left[\cos(\xi)\left(|\tilde{1}\rangle+|\tilde{2}\rangle\right)/\sqrt{2}
-\sin(\xi)|3\rangle\right]e^{-i\kappa' z},
\end{eqnarray}
where $\kappa'\equiv k^{l}_{r}[1-\cos(\xi)]$,
$|\tilde{1}\rangle\equiv |1\rangle e^{ik^{l}_{r}(x+z)}$, and
$|\tilde{2}\rangle\equiv |2\rangle e^{-ik^{l}_{r}(x-z)}$. Clearly,
within the dark-state subspace the dynamical phase is always zero.
So we focus on the time evolution in this dark subspace. Any state
therein can be expanded as $c_{1}|D_{1}\rangle+c_{2}|D_{2}\rangle$.
This expansion hence defines a dark state representation. In this
representation, the mechanical momentum operator becomes
\cite{Juzeliunas2008PRL, groupeqnote}
\begin{eqnarray}
\mathbf{P}^{D}=-i\hbar\tilde{\nabla}+ \hbar \kappa
(\sigma_{x}\hat{e}_{x}+\sigma_{z}\hat{e}_{z}), \label{nonabelianP}
\label{Pgroup}
\end{eqnarray}
where $\sigma_{x,z}$ are Pauli matrices,
$\kappa=\cos(\xi)k_{r}^{l}$, and $\tilde{\nabla}$ represents the
gradient in the dark-state representation.  As in Ref.
\cite{Juzeliunas2008PRL}, we also introduce an additional constant
shift to state $|3\rangle$, denoted by a matrix operator $V_{s}$,
such that the total effective Hamiltonian becomes
$H^{\text{eff}}_{D}=\frac{(\mathbf{P}^{D})^{2}}{2m}. $
The eigenstates of $H^{\text{eff}}_{D}$ are
\begin{equation} \label{Weigen}
|\Psi^{D,\pm}\rangle=\frac{1}{2}\left(\begin{array}{c}
1\mp ie^{i\varphi_{k}}\\
-i\pm e^{i\varphi_{k}}
\end{array}
 \right)e^{i\mathbf{k}\cdot\mathbf{R}},
\end{equation}
where $\varphi_{k}$ is the angle between the $x$-axis and the atom
wavevector $\mathbf{k}$ in Eq. (\ref{Weigen}).  According to Eq.
(\ref{nonabelianP}), it is straightforward to show that the
eigenstates $|\Psi^{D,\pm}\rangle$ have the momentum
$(\mathbf{k}\pm\kappa \hat{\mathbf{k}})\hbar$.

Consider then two simple scenarios of laser field manipulation. In
the first scenario, we adiabatically move all the laser beams along
the negative $z$ axis. The non-Abelian geometric phases thus induced
are determined by
\begin{eqnarray} \label{displace}
i\frac{d}{d z}\left(\begin{array}{c}
c_{1}\\
c_{2}
\end{array}
 \right)
& =&-\left(\begin{array}{cc}
i\langle D_{1}|\frac{\partial}{\partial z}|D_{1}\rangle & i\langle D_{1}|\frac{\partial}{\partial z}|D_{2}\rangle\\
i\langle D_{2}|\frac{\partial}{\partial z}|D_{1}\rangle & i\langle
D_{2}|\frac{\partial}{\partial z}|D_{2}\rangle
\end{array}
 \right)
  \left(\begin{array}{c}
c_{1}\\
c_{2}
\end{array}
 \right) \nonumber \\
& =&\left(\begin{array}{cc}
\kappa & 0\\
0 & -\kappa
\end{array}
 \right)
  \left(\begin{array}{c}
c_{1}\\
c_{2}
\end{array}
 \right).
\end{eqnarray}
Because the matrix on the right side of Eq. (\ref{displace}) turns
out to be independent of ${\bf R}$, one may naively conclude that
the non-Abelian geometric phases induced here will not affect matter
wave propagation.  But this is incorrect. To see this let us assume
the initial state to be the $|\Psi^{D,-}\rangle$ state in Eq.
(\ref{Weigen}) with $\varphi_k=0$. This initial state has a
mechanical momentum $\hbar(k-\kappa)$ along the $x$-axis, where
$k\equiv |{\bf k}|$.   Let the final state, a function of the
displacement $d_z$ along $-z$,   be $|\Psi^{D}_{f}(d_z)\rangle$. It
is enlightening to consider a specific case, e.g.,
$d_z=\frac{\pi}{4\kappa}$.  Neglecting matter wave propagation
during the period of geometric phase imprinting,  Eq.
(\ref{displace}) then gives
\begin{eqnarray}
|\Psi^{D}_{f}(\pi/4\kappa)\rangle=\left[\frac{1+i}{2\sqrt{2}}\left(\begin{array}{c}
1\\
1
\end{array}
 \right)- \frac{1-i}{2\sqrt{2}}\left(\begin{array}{c}
1\\
-1
\end{array}
 \right) \right]e^{ikx}.
 \label{Psif}
\end{eqnarray} The two components on the right side of
Eq. (\ref{Psif}) represent a superposition of the two eigenstates
$|\Psi^{D,+}\rangle$ and $|\Psi^{D,-}\rangle$ in Eq. (\ref{Weigen}).
Because these two eigenstates possess different mechanical momenta $
(k+\kappa)\hbar$ and $(k-\kappa)\hbar$, we thus have the result that
the non-Abelian geometric phases generated here can split matter
waves.

\begin{figure}[t]
\begin{center}
\vspace*{-0.cm}
\par
\resizebox *{6.5cm}{6.5cm}{\includegraphics*{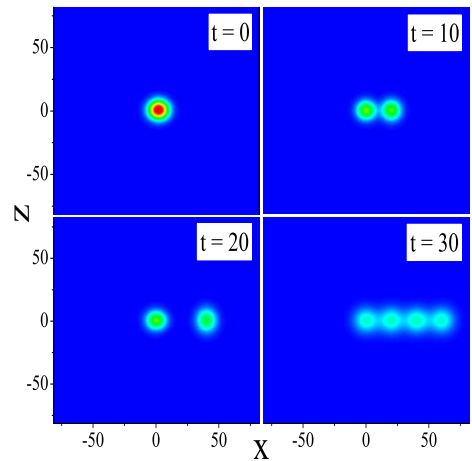}}
\end{center}
\par
\vspace*{-0.5cm} \caption{(Color Online) Numerical simulation of
wavepacket splitting induced by imprinting non-Abelian geometric
phases. Here $t$ is in units of $m/(\hbar \kappa^{2})$; $x$ and $z$
are in units of $1/\kappa$. See the text for details.}\label{fig1}
\end{figure}

This result is also verified by our numerical simulations based on
the full Hamiltonian
$-\frac{\hbar^{2}\nabla^{2}}{2m}+H_{\text{RWA},4}+V_{s}$. In
particular, in our simulations the initial state is chosen as a
Gaussian wavepacket instead of a plane wave considered in Eq.
(\ref{Weigen}). As an example we choose $\Omega_0/\hbar=1200
\hbar\kappa^2/m$.  The duration of the imprinting process is chosen
to be $0.063 m/(\hbar \kappa^{2})$, (In real units, for $m=10^{-25}$
Kg, $\kappa\sim 10^{6}$ m$^{-1}$, the duration is $\sim 60 \ \mu$S
and $\Omega_0/\hbar\sim 10^6$ Hz. The cold-atom version of the
Zitterbewegung oscillation \cite{clark} is negligible for these
parameters). Figure 1 depicts the simulation results for
$\varphi_k=0$, $k=\kappa$. At $t=0$ we have initiated the first
non-Abelian geometric phase imprinting for $d_z=\pi/(4\kappa)$. As
seen in Figs. 1(b)-(c),  an initial wavepacket of zero group
velocity indeed splits into two parts: one part holds a zero group
velocity and the other part moves along the $x$ axis at an average
mechanical momentum of $2\hbar \kappa$.
Then, at $t=20$, we move all the lasers up by $d_z=\pi/(4\kappa)$ to
induce a second imprinting process. Figure 3(d) shows that this
further splits the wavepacket into four copies of equal weights, a
result also derivable from Eq. (\ref{displace}). Choosing a
different $d_z$ at $t=20$ for the second imprinting process, we may
even exchange the momenta between the two separated sub-wavepackets.
Extending the strategy here, in principle one can split a wavepacket
into $2^n$ copies with an arbitrary integer $n$. This should be
useful for atom optics applications such as atom interferometry.
Other simulation results (not shown) also indicate that so long as
$\Omega_0$ is sufficiently large, then on one hand the matter wave
propagation during the adiabatic processes can be made negligible
and on the other hand the adiabaticity in the internal motion is
well maintained.


In the second scenario, we slowly rotate all the three laser beams
clockwise in the $x$-$z$ plane. The non-Abelian geomertic phase at
the local point $(R,\theta)$ is determined by
\begin{eqnarray} \label{Rodisplace}
&i\frac{d}{d \gamma}\left(\begin{array}{c}
c_{1}\\
c_{2}
\end{array}
 \right)
=-\left(\begin{array}{cc}
i\langle D_{1}|\frac{\partial}{\partial \gamma}|D_{1}\rangle & i\langle D_{1}|\frac{\partial}{\partial \gamma}|D_{2}\rangle\\
i\langle D_{2}|\frac{\partial}{\partial \gamma}|D_{1}\rangle &
i\langle D_{2}|\frac{\partial}{\partial \gamma}|D_{2}\rangle
\end{array}
 \right)
  \left(\begin{array}{c}
c_{1}\\
c_{2}
\end{array}
 \right) \nonumber \\
& =\kappa R \left(\begin{array}{cc}
\cos(\theta+\gamma) & -\sin(\theta+\gamma)\\
-\sin(\theta+\gamma) & -\cos(\theta+\gamma)
\end{array}
 \right)
  \left(\begin{array}{c}
c_{1}\\
c_{2}
\end{array}
 \right),
\end{eqnarray}
where $0\leq \gamma\leq \alpha $ is the angle of rotation that
serves as the adiabatic parameter. An analytical solution to Eq.
(\ref{Rodisplace}) is not found and hence we solve it
computationally.  After obtaining the numerical non-Abelian
geometric phases for each point $(R,\theta)$ from Eq.
(\ref{Rodisplace}), we imprint the spatially dependent geometric
phases onto the system at $t=0$ and then let it evolve under
$H_{D}^{\text{eff}}=\frac{{\bf P}_{D}^2}{2m}$ (also with the rotated
laser fields).  For the example shown in Fig. 2, the initial state
is the state $|\Psi^{D,-}\rangle$ in Eq. (\ref{Weigen}) multiplied
by a Gaussian profile, with $k=\kappa$ and $\varphi_k=-\pi/2$. This
initial state hence has a zero group velocity according to Eqs.
({\ref{Pgroup}) and (\ref{Weigen}).
As shown in Fig. 2 for $\alpha=3\pi/2$, the phase imprinting
substantially impacts the ensuing evolution.  The evolving
wavepacket first displays interesting interference patterns,
followed by a certain degree of self-focusing,  and later it starts
to display some moon-like patterns. As expected, the patterns in
real space at later times begin to resemble the $t=0$ wavepacket
pattern plotted as a function of $P^{D}_{x}$ and $P^{D}_{z}$,
namely, the mechanical momenta along the $x$-axis and $z$-axis.
Remarkably, even if we let $\alpha=2\pi$, i.e., with all laser
fields returning to their original configuration, similar results
can still be obtained. We have checked that essentially the same
results can be obtained using the full Hamiltonian
$-\frac{\hbar^{2}\nabla^{2}}{2m}+H_{\text{RWA},4}+V_{s}$. However,
we find that in order to fully neglect matter wave propagation
during the imprinting, the duration of the adiabatic process here
should be about two orders of magnitude shorter than in the first
non-Abelian example.  Hence $\Omega_0$ must be about two orders of
larger.

\begin{figure}[t]
\begin{center}
\vspace*{-0.cm}
\par
\resizebox *{9cm}{7.6cm}{\includegraphics*{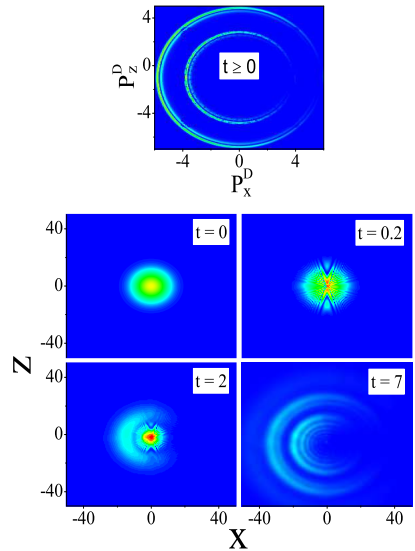}}
\end{center}
\par
\vspace*{-0.5cm} \caption{(Color Online) Simulation of the
wavepacket dynamics after imprinting non-Abelian geometric phases
via adiabatic rotation of the three laser beams in a tripod scheme.
The wavepacket plotted in momentum space (top panel, in units of
$\hbar\kappa$) will remain the same for any $t \geq 0$. Length units
(for the bottom four panels in $x$-$z$ space) are the same as in
Fig. 1.} \label{fig2}
\end{figure}

In summary,  we have demonstrated the concept of all-optical
imprinting of both Abelian and non-Abelian geometric phases onto
matter waves. The adiabatic paths considered here are among the
simplest. Exploring more complicated adiabatic paths might offer
extensive laser control of matter wave propagation. Extending this
work to the imprinting of non-adiabatic geometric phases is also of
considerable interest.

This work was supported by WBS grant No. R-710-000-008-271 (ZQ and
CH) under the project ``Topological Quantum Computation",  and by
the NUS ``YIA" fund (WBS grant No.: R-144-000-195-123) (JG).

\bibliography{/home/bwu/references/berry}

\begin{thebibliography}{99}
\bibitem{footnote}The last equality in Eq. (\ref{GP}) will become
mathematically exact if $-\Delta_{\text{min}}=\Delta_{\text{max}}$
or if $|\Delta_{\text{min}}/\Omega_0|\rightarrow \infty $ and
$|\Delta_{\text{max}}/|\Omega_0|\rightarrow \infty$.
\end{thebibliography}
\bibliographystyle{apsrev}

\end{document}